\documentclass[12pt]{article}
\usepackage{epsfig,rotating}
\begin{document}
\baselineskip 16pt plus 2pt minus 2pt

\newcommand{\beq}{\begin{equation}}
\newcommand{\eeq}{\end{equation}}
\newcommand{\beqa}{\begin{eqnarray}}
\newcommand{\eeqa}{\end{eqnarray}}
\newcommand{\dida}[1]{/ \!\!\! #1}
\renewcommand{\Im}{\mbox{\sl{Im}}}
\renewcommand{\Re}{\mbox{\sl{Re}}}
\newcommand{\PRD}[3]{{Phys.~Rev.} \textbf{D#1},({#2}) #3}
\newcommand{\PLB}[3]{{Phys.~Lett.} \textbf{B#1}, ({#2}) #3}
\newcommand{\PRL}[3]{{Phys.~Rev.~Lett.} \textbf{#1}, ({#2}) #3}
\newcommand{\NPB}[3]{{Nucl.~Phys.} \textbf{B#1}, ({#2}) #3}
\newcommand{\NPA}[3]{{Nucl.~Phys.} \textbf{A#1}, ({#2}) #3}
\def\simge{\hspace*{0.2em}\raisebox{0.5ex}{$>$}
     \hspace{-0.8em}\raisebox{-0.3em}{$\sim$}\hspace*{0.2em}}
\def\simle{\hspace*{0.2em}\raisebox{0.5ex}{$<$}
     \hspace{-0.8em}\raisebox{-0.3em}{$\sim$}\hspace*{0.2em}}

\begin{titlepage}

%\hfill{Version 05/29/00 pm}

%\hfill{TRI-PP-00-XX}

\vspace{1.0cm}

\begin{center}
{\large {\bf  Bulk Higgs Boson Decays in Brane Localized Gravity }}\\

\vspace{1.2cm}

Chia-Hung V. Chang$^{\ast}$\footnote{email:
chchang@phy03.phy.ntnu.edu.tw}
and
J. N. Ng$^{\ast \ast ,\dagger}$\footnote{email: misery@triumf.ca}

\vspace{0.8cm}
$^{\ast}$ Physics Department, National Taiwan Normal University, Taipei,
Taiwan\\
$^{\ast \ast}$TRIUMF, 4004 Wesbrook Mall, Vancouver, BC, Canada V6T 2A3\\
$^{\dagger}$National Center for Theoretical Sciences, P.O. Box 2-131,
Hsinchu, Taiwan 300\\[.4cm]
\end{center}

\vspace{1cm}

\begin{abstract}
We embed the Standard Model in the Randall-Sundrum model of
5 dimensional brane localized gravity. 
The SM gauge and chiral fermion fields are restricted on the 4D visible
brane whereas the Higgs and the right-handed neutrino 
are assumed to be 5D bulk fields. We calculate the
effective couplings of the lowest mass Higgs field to the SM fermions and 
to the gauge bosons and find that
the couplings are enhanced. 
Furthermore, the invisible decay width of a bulk Higgs of mass 150 GeV 
is shown to be large. 

\end{abstract}  

\vspace{2cm}
\vfill
\end{titlepage}
%\section{Introduction}
	It has been  a long held belief in  particle physics that the
gravitational force is too feeble, i.\ e.\  the Planck scale too high, to have
an  impact on physics  at  the  weak   scale.   Recent  theoretical
developments have shed new  light on this  sixteen order  of magnitude
disparity  between the electroweak  scale and  the Planck scale, which is one of 
several hierarchy problems in particle physics. To this end, authors made ingenious  
uses of  extra spatial
dimensions which are present in  many extensions of the Standard Model
(SM) such  as string theories  and supergravity models. The  number of
extra dimensions  $n$ can range  from 1 to  6 or 7.  Another important
ingredient  is  to confine  the  SM chiral  fermions  on  one or  more
3-branes which  are stable  topological objects in string  theory. From
the four dimensional field theory  point of view these chiral fermions
have only the usual  Minkowski spacetime dependence given by $x^{\mu}$
where  $\mu=0,1,2,3$   and  not  on  the  coordinates   of  the  extra
dimensions. On the  other hand, gravity is allowed  to spread into the
extra  dimensions. This opens  up novel  settings where  new interplay
between electroweak and gravitational physics can take place.

	There are two main constructions to resolve the hierarchy problem
under the general framework described above. The first one is to assume
that the geometry of spacetime is factorizable and given by $M^4 \times S^n$ where $M^4$ denotes
the usual four dimensional Minkowski space and the geometry of the extra dimensions is usually taken
to be a $n$-tori for simplicity. The four dimensional Planck mass
$M_P$ is related to the fundamental scale of the higher dimensnional theory
$M_*$ by the relation \cite{ADD}
\beq
\label{scal}
 M_P^2=M_*^{n+2}(2\pi R_1)(2\pi R_2)\ldots (2\pi R_n) \;
\eeq
where $R_i$ with $i=1,2,\ldots,n$ denotes the compactification radii. For 
simplicity one takes all the radii to be equal to $R$. A very general prediction of 
this scenario is a modification of the Newtonian gravitational law which is well measured at large
distances. From this one concludes that $n\geq 2$ and the gravitational law 
is only changed at short distances below the micron range \cite{ADD}. For $n=2$
astrophysical considerations set the limit on $M_\ast \geq 50 \mathrm TeV$ and
$R \leq 0.3 \mu$m \cite{asc}. As $n$ increases the constraints becomes less stringent.    

        An alternative scenario is given by Randall and Sundrum \cite{RS1}. In the simplest version
spacetime is taken to be five dimensional with the fifth dimension, $y$, compactified on a $S_1/Z_2$
 orbifold of radius $r_c$. Hence, we can write $y=r_c\phi$ and $-\pi \leq \phi \leq \pi$. The points $(x,\phi)$
and $(x,-\phi)$ are identified. Two 3-branes with equal and opposite tensions are located at the orbifold 
fixed points: a visible brane at 
$\phi=\pi$ where all the SM particles are confined and a hidden brane at $\phi=0$ where gravity is localized. 
The metric that solves 
the Einstein equations is given by \cite{RS1}
\beq
\label{warp}
 ds^2=e^{-2kr_c|\phi|}\eta_{\mu\nu}dx^{\mu}dx^{\nu} - r_c^2 d\phi^2 \;
\eeq
where $k$ is a parameter of the order of the fundamental mass scale $M$ of the 5D theory and
$\eta_{\mu \nu}$ is the Minkowski flat metric with the signature $(+---)$. The exponential 
factor in Eq.(\ref{warp}) is known as 
the warp or conformal factor. The 4D Planck scale is calculated to be given by 
\beq
\label{Pl}
 M_P^2= \frac{M^3}{k}[1-e^{-2kr_c\pi}]   \;
\eeq
and hence is of the order of the scale $M$. Furthermore, any field confined on the visible brane
at $\phi=\pi$ with mass $m_0$ will be rescaled to have a physical mass given by $m_0e^{-kr_c\pi}$.
With the value of $kr_c=\mathrm 12$ a weak scale is dynamically generated with all fundamental masses
of the order of $M_P$.

        Treating Eq.(\ref{warp}) as a background metric, particles  of different spins represented by full
fledged bulk fields have been studied. The scalar field was treated in \cite {GW} and it was found that 
many of its properties 
are controlled by the warp factor. In addition it can be used to stabilize the extra dimension \cite{sed}.
 The SM singlet fermion is studied in \cite {GN} as a means of generating
a small neutrino mass without using the seasaw mechanism.
 Issues of bulk gauge fields are examined in \cite {Bgf} and the embedding the SM
in the full 5D bulk is given in \cite {CHN}, \cite {DHR} and \cite {DHRnew} . A characteristic of this scenario has emerged from
 these studies. In all cases the zero modes and
the Kaluza-Klein (KK) excitations of the bulk fields are given by the roots of Bessel functions of different
orders relating to their intrinsic spin. This is in sharp contrast to the case of factorizable geometry
where the masses of the KK excitations are typically
$m^2=\sum_{i}\,n_i^2 /R_i^2$ where $n_i$ are
integers.
It was also noted in \cite {CHN} that a fine tuning problem will emerge
with the SM in the bulk if the Higgs boson is also allowed to extend into the fifth dimension.
  
        In this paper we study a model with the SM chiral fermions and the gauge bosons all confined to
a 3-brane at $\phi=\pi$ but the Higgs boson is taken to be a bulk field. This is similar to the bulk
scalar field studied in
\cite {GW} but the bulk Higgs field develops a non-zero VEV 
after spontaneous symmetry breaking. The motivation here is
to keep the feature that a bulk scalar field can confine fermions on a kink as found in \cite {kin}. Since we 
are confining the gauge fields on the brane, 
the Higgs field has local gauge symmetry on the visible brane.
However, viewed in the fifth dimension, this is a global symmetry. Specifically, the brane fermions,
$\psi(x)$, a $U(1)$ brane gauge field $A_{\mu}(x)$ and the bulk Higgs field, $H$, transform respectively as
follow:
\beqa
\label{gtran}
\global\def\theequation{4a}
\psi(x)    & \rightarrow & e^{i\Lambda (x)}\psi(x)\\
\global\def\theequation{4b}
A_{\mu}(x) & \rightarrow & A_{\mu}(x) + \partial_{\mu} \Lambda (x) \\
\global\def\theequation{4c}
H(x,\phi)  & \rightarrow & e^{i\Lambda (x)} H(x,\phi).       \;
\eeqa
where the gauge function $\Lambda$ depends on $x^\mu$ only. We also include a bulk singlet fermion
field denoted by $\Psi (x^\mu,\phi)$ which will serve as a right-handed neutrino,
like in \cite{GN}.  
The focus of our paper will be the decay modes of a physical
Higgs of mass $M_H$ between  125 and 250 GeV. With a mass in this range,
the Higgs boson predominantly 
decays into a pair of b-quarks or gauge bosons. 
We shall see in detail later that this is also true for a bulk Higgs 
boson. Hence we  can examine quantitatively how
the difference between the bulk Higgs and the SM Higgs boson 
can be manifested in the ongoing and future
Higgs boson searches at high energy colliders.

        The relevant action for the model described above can be written in four separate pieces. We begin with
 the bulk Higgs field action in 5D and it is given by:

\global\def\theequation{\arabic{equation}}
\setcounter{equation}{5}
\beq
\label{bha}
 S_H = \int d^4 x \int_{-\pi}^{\pi} d\phi \sqrt{G}\left[ G^{AB}D_A H_0^{\dagger} D_B H_0 -
\frac{\lambda}{4M}\left(
H_0^{\dagger}H_0-\frac{v_0^3}{2}\right)^2 \right]
\eeq
where $H_0$ denotes the bulk Higgs field, which is a weak doublet.
 $G$ is the determinant of the metric tensor $G_{AB}$
of Eq.(\ref{warp}). $\lambda$ is a dimensionless parameter and 
 $v_0$ is the scale that characterizes the 5D vacuum expectation value of $H_0$. 
 We use the notation that 
the capital Roman letters $A,B,...$ denote 5D coordinates, the lower case letters $a,b,..$ denote its tangent space
coordinates, and the Greek letters $\mu,\nu,...$ label Minkowski 
space coordinates. $v_0$  is expected to be of order $M$. 
The gauge covariant derivative $D_A$ is given by $D_{\mu}=\partial_{\mu} + igA_{\mu}$ for the Minkowski coordinates
where $A_{\mu}$ is the 4D brane  gauge field
and $D_5=\partial_5$ for the fifth dimension.

        We begin by studying the scalar sector.  The real  charge zero component of the doublet 
develops a non-zero VEV $v_0^{\frac{3}{2}}/\sqrt{2}$. After shifting the
scalar field by $H_0 \rightarrow (H+v_0^{\frac{3}{2}})\,/\sqrt{2}$,
 we obtain an action similar to that given in \cite{GW}.
To obtain the masses of the bulk Higgs boson and its KK excitations we first substitute the metric $G_{AB}$ into
Eq.(\ref{bha})
 and use the Kaluza-Klein decomposition of $H$:
\beq
\label{kkh}
H(x,\phi)=\frac{1}{\sqrt {r_c}}\sum_{n} h_n(x)y_n(\phi). \;
\eeq
Define that $\sigma \equiv kr_c\phi$.
 If the $y_n(\phi)$ is chosen to satisfy the normalization condition:
\beq
\label{hnorm}
\int_{-\pi}^{\pi}d\phi \;e^{-2\sigma}y_m(\phi)y_n(\phi)=\delta_{mn}
\eeq
and the mass eigenvalue equation
\beq
\label{heig}
-\frac{1}{r_c^2}\frac{d}{d\phi}\left( e^{-4\sigma}\frac{dy_n}{d\phi}\right) + m^2 e^{-4\sigma}y_n =
m_n^2e^{-2\sigma}y
\eeq
with $m^2= \lambda v_0^3/M$, the effective 4D action then simplifies to:
\beq
\label{4dh}
S_h^{(4)}=\frac{1}{2}\sum_{n}\int d^4x \; [\eta^{\mu\nu}\partial_{\mu} h_n \partial _{\nu}h_n -m_n^2 h_n^2]
\eeq
which identify $h_n$ as the $n^{\mathrm th}$ KK Higgs excitation with mass $m_n$ given by Eq. (\ref{heig}).
The lowest mass state will be the one we are interested in. It is useful to introduce the variables:
\beq
\label{vari}
f_n \equiv e^{-2\sigma}y_n , \hspace{1.0cm} z_n \equiv \frac{m_ne^{\sigma}}{k} ,\hspace{1.0cm} 
{\mathrm {and}}\hspace{1.0cm} \omega\equiv \frac{m}{k}\;
\eeq
Then  Eq. (\ref{heig}) can be 
cast into the standard form:
\beq
\label{hbess}
z_n^2\frac{d^2f_n}{dz_n^2} + z_n\frac{df_n}{dz_n} + \left[z_n^2-(4+\omega ^2)\right] f_n = 0.
\eeq
The solutions of this equation are  Bessel functions of order $\nu=\sqrt{4+\omega^2}$: $J_{\nu}(z_n)$. 
It is clear that $\nu \geq 2$ with the lower bound  given by $\lambda=0$ which would be obtained by a higher
dimension radiative symmetry breaking mechanism.
After imposing the boundary conditions that the derivative of $y_n$ be continuous at $\phi=0,\pi$ and the approximation 
$e^{kr_c\pi}\gg 1$ and we obtain the following equation:
\beq
\label{hbc}
          x_{n\nu}J_{\nu-1}(x_{n\nu})=(\nu -2)J_{\nu}(x_{n\nu})
\eeq
which gives the eigenvalues of $x_{n\nu} \equiv \frac{m_n}{k}e^{kr_c\pi}$. For the case of $\omega \simeq 0$
the eigenvalues are essentially the  roots of $J_1(x)$. Numerically the first two values are 3.83 and 7.02. The 
lowest eigenvalues of $x_1$ vary between 3.8 to 12.5 when $\omega = \mathrm {0.1} \sim \mathrm { 10} $. 
We identify the lowest mode to be the lightest Higgs boson of mass $m_1$; then
\beq
\label{hscon}
k\epsilon\equiv ke^{-kr_c\pi}= \frac{m_1}{x_1}.
\eeq

In Table 1 we give $k\epsilon$ for different vaules of $\omega$ and $m_1$. The masses of  
the first KK excited states are also shown. It can be seen that in this
scenario we are led to a tower of closely spaced Higgs states for a wide range of $\omega$ values.

        Now we continue with the gauge boson-Higgs interaction contained in Eq.(\ref{bha}). Using standard notations 
the action for the bulk Higgs-$W$ interaction is given by 
\beq
\label{gha}
 \int d^4 x \int_{-\pi}^{\pi} d\phi \sqrt{G}\left[ G^{\mu\nu}W_{\mu}^{-} H_0^{\dagger} W_{\nu}^{+} H_0
\right]
\eeq
After spontaneous symmetry breaking, the $W$ boson mass is generated via:
\beq
 \frac{1}{4}\int d^4 x \, r_c e^{-4kr_c \pi} e^{2kr_c\pi} g^2 \left[ \eta^{\mu\nu}
  W_{\mu}^{-}  W_{\nu}^{+} v_0^3 \right]
\eeq
From this we can relate the physical $W$ mass, $M_W$, to the 5D parameters; i.e.
\beq
\label{wmass}
M_W = \frac{1}{2} g v_0 e^{-kr_c \pi} \sqrt{r_c v_0}
\eeq
or equivalently we can write the Fermi scale $v$ as
\beq
\label{fermi}
 v =\epsilon v_0\sqrt{r_cv_0}=250 \:\mathrm {GeV}.
\eeq
We can now calculate the coupling of the KK Higgs to $W^{+}W^{-}$. Using Eqs.(\ref{gha}),(\ref{kkh}),
(\ref{vari}) and noting that the bulk scalar fields are evaluated at the visible brane, we find
the $h_n W^{+}W^{-}$ coupling is
\beq
\label{hww}
gM_W \sqrt{\frac{k r_c}{1-(\frac{\omega}{x_{n\nu}})^2}}
\eeq
Compare this with the case in SM where the coupling is $gM_W$. Similar conclusion applies
to the Higgs-$Z$ boson coupling. Hence, in this model the ratio
of the width of the lowest Higgs state decaying into 2 gauge bosons 
compared to that of the SM Higgs decay is
\beq
\label{rg}
R_g=\frac{\Gamma (h_1\rightarrow W^{+}W^{-},ZZ)}{\Gamma (H_{SM}\rightarrow W^{+}W^{-},ZZ)}=\frac{kr_c}{1-(\frac{\omega}
{x_{1\nu}})^2 }.
\eeq
We note that $x_1$ is determined by the order of the Bessel equation
and hence  on $\omega$ but not on the choice of 
$m_1$; then $R_g$ takes the values 12 and 33.3 for $\omega= \mathrm {0.1, 10}$ respectively and it is most sensitive
to the choice of $kr_c$.

         Next we introduce the brane fermions located at $\phi=\pi$ and their interactions with $H$. 
Our discussions will be given in terms of the SM lepton doublet $L_0$ and the right-handed $e_{0R}$ and
the subscript $0$ is used to denote unrenormalized fields. The results obtained can be easily carried 
over to the quarks. 
Using the notation that $g_{\mu \nu}^{vis}(x_{\mu}) = G_{\mu \nu}(x_{\mu},\phi=\pi)$ 
and $g^{vis}$ its determinant,
the action is given by
\beq
\label{bferm}
 S_{bf}=\int d^4x\; \sqrt{-g^{vis}} \left( \bar{L}_0\hat{\gamma}^{\mu}\partial_{\mu} L_0 + \bar{e}_{0R}
\hat{\gamma}^{\mu}e_{0R} \right ) -\frac{\hat{y}_e}{\sqrt{M}}\int d^4x\! \sqrt {-g^{vis}}\bar{L}_0H_0
e_{0R}+h.c. 
\eeq
where 
\beqa
\label{5dg}
 \hat{\gamma}^{\mu} &=& E_a^{\mu}(\phi=\pi)\gamma^a \nonumber   \\
                    &=& e^{kr_c\pi}\gamma^{\mu}.
\eeqa
In Eq.(\ref{bferm}) $H_0$ is at $\phi=\pi$ and in Eq.(\ref{5dg}) the inverse vielbein, 
$E^A_a = \mathrm {diag}(e^{\sigma},e^{\sigma},e^{\sigma},e^{\sigma},\frac{1}{r_c})$. 
A ubiquitous Yukawa coupling, $\hat{y}_e$, is introduced in Eq.(\ref{bferm}) and is a free parameter. 
The SM chiral fermions reside on the brane. The field $H$ will be expanded
via Eq.(\ref{kkh}) and evaluated at $\phi=\pi$. The kinetic term will require a rescaling due to 
nontrivial $g^{vis}$ and $E^A_a$.
The brane fermion wavefunction rescaling is
\beq
\label{bfsc}
 L_0 = e^{\frac{3}{2}kr_c\pi} L
\eeq 
and the kinetic term for $L$ is in the canonical form. The $e_{0R}$ field  is similarly rescaled. Spontaneous 
symmetry breaking will generate a mass for the electron via the Yukawa term in Eq.(\ref{bferm}). This is
given by 
\beq
\label{emass}
m_e=\hat{y}_e \epsilon v_0 \sqrt{\frac{v_0}{2M}}
\eeq
A second product of this manipulation is an expression for  the 
effective Yukawa coupling  of the interaction 
$\bar{e}_Le_Rh_n$ on the visible brane and it is
\beq
\label{hee}
 y_e^{eff}=\frac{gm_e}{2M_W}\sqrt{\frac{kr_c}{ 1-(\frac{\omega}{x_{n\nu}})^2 }} ,
\eeq
where we have used  Eqs.(\ref{vari}), (\ref{hbc}) and (\ref{wmass}).
Barring fortuitous tuning of parameters it is clear that the decay width of the lowest state bulk Higgs, $h_1$, 
into a fermion pair will be different 
from the SM Higgs boson due to the square root factor in Eq.(\ref{hee}). 
Indeed the ratio of the width of $h_1$ decaying to a fermion pair to that of a SM Higgs of the same
mass is predicted to be
\beq
\label{hffbr}
R_f=\frac{\Gamma (h_1\rightarrow \bar{f}f)}{\Gamma (H_{SM}\rightarrow \bar{f}f)}=\frac{kr_c}
 { 1-(\frac{\omega}{x_{1\nu}})^2 }
\eeq
Noticed that the gauge boson width and all the fermion widths are enhanced by the same factor:
$R_f=R_g$.

        This is a convenient point to pause and take stock of the parameters of the model. Aside 
from the Yukawa coupling there are five 5D parameters, {\em viz.} $v_0, M, k, r_c,$ and $m$, which
replaces $\lambda$. As argued in \cite{RS1} that $kr_c$ is of O(10) in order to solve the hierarchy problem.
Without loss of generality we take $kr_c=$ 12 as a benchmark value. 
We have also seen that $\omega = m/k$
is of order unity and this is a parameter we vary as was done in Table 1. Then by choosing a value for $m_1$ 
the value of $k$ and $v_0$ is determined by Eq.(\ref{hbc}) and (\ref{fermi}).
Similarly the fundamental 5D scale $M$ will be fixed by Eq.(\ref{Pl}).

        As noted in \cite{GN} that a SM singlet bulk fermion, $\Psi$, can be added to the 
brane localized gravity model and generate a small Dirac neutrino mass for $\nu_{eL}$ as an
alternative to the seesaw mechanism. The construction 
of the action for $\Psi$ is given in \cite{GN} and we generalized it to include a bulk Higgs mechanism. 
The action of the bulk fermion takes the form \cite {EGH}
\beq
\label{bufm}
S_{\Psi}=\int d^4x \int_{-\pi}^{\pi}d\phi \sqrt{G}\biggl\{ E^A_a\left[ \frac{i}{2}\bar{\Psi}\gamma^a
(\partial_A-\overleftarrow{\partial}_A) \Psi +\frac{\omega_{bcA}}{8}\bar{\Psi}\{\gamma^a,\sigma^{bc}\}\Psi \right]
-m_D \,{\rm sgn}(\phi)\bar{\Psi}\Psi \biggr\}
\eeq
where we have included a bare Dirac mass $m_D$ and $\omega_{bcA}$ is the spin connection 
associated with the warp metric and it gives no contribution to
the physics we are discussing. A convenient choice of the Dirac matrices is $\gamma^a = (\gamma ^{\mu},
i\gamma^5)$. In 5D one can define the left- and right-handed spinors $\Psi_{L,R} = \frac{1\mp \gamma^5}{2} \Psi$.
We are mainly interested in $\Psi_R$ and will present enough formulas so that we are self contained and 
 our notations are clear.
Details of their derivations can be obtained in \cite {GN}. After using the KK decompositions of $\Psi_{L,R}$ is
\beq
\label{kkpsi}
\Psi_{L,R}=\frac{1}{\sqrt{r_c}}\sum_ne^{2\sigma}\psi_{nL,R}(x)\hat{f}_{nL,R}(\phi)
\eeq
The functions $\hat{f}_{nL,R}$ are separate complete sets of orthonormal functions. 
After imposing the ${\cal {Z}}_2$ orbifold symmetry and the periodic boundary condition of $\Psi_{L,R}(x,\pi) =
\Psi_{L,R}(x,-\pi)$ the range of $\phi$ is the interval  $[0, \pi]$, and we get the 4D action for the bulk neutrino
and its KK excitation 
\beq
\label{4dbf}
S_{\psi}= \sum_n \int d^4x (\bar{\psi}_n i\gamma^{\mu}\partial_{\mu}\psi_{n} - \mu_n \bar{\psi_n}\psi_n)
\eeq 
where $\psi=\psi_L+\psi_R$ and $\mu_n \geq 0$. As in the scalar case 
the functions $\hat{f}_{nL,R}$ satisfy
\beq
\label{ffn}
\int_0^{\pi} d\phi e^{\sigma}\hat{f}^*_{nR}\hat{f}_{mR}=\int_0^{\pi} d\phi e^{\sigma}\hat{f}^*_{nL}\hat{f}_{mL}
=\delta_{nm}  
\eeq
and the equation
\beq
\label{feig}
\left( \pm \frac{1}{r_c}\partial_{\phi}-m_D \right) \hat{f}_{nL,R} = -\mu_ne^{\sigma}\hat{f}_{nR,L}.
\eeq
With the following change of variables:
\beq
\label{newvari}
t\equiv \epsilon e^{\sigma}, \hspace{1cm} \tilde{\nu}\equiv \frac{m_D}{k}, \hspace{1cm} {\mbox {and}}\hspace{1cm}
\tilde{x}_n\equiv \frac{\mu_n}{\epsilon k}
\eeq
and 
\beqa
\label{newfv}
\global\def\theequation{32a}
{f}_{nL,R}(t)&=&\frac{\hat{f}_{nL,R}(\phi)}{\sqrt{kr_c\epsilon}},\\
\global\def\theequation{32b}
        \tilde{f}_{nL,R}(t)&=& \frac{f_{nL,R}(t)}{\sqrt{t}} \;
\eeqa 
we can combine the two first order equations into the standard Bessel equation of half integer order:
\beq
\global\def\theequation{\arabic{equation}}
\setcounter{equation}{33}
\label{fbess}
 t_n^2\frac{d^2\tilde{f}_{n,L,R}}{dt_n^2}+t_n\frac{d\tilde{f}_{nL,R}}{dt_n}+\left[ t_n^2-(\tilde{\nu}\mp
\frac{1}{2})\right] \tilde{f}_{nL,R}=0
\eeq
where $t_n\equiv \tilde{x}_nt$.
In particular we are interested in the zero mode with $\mu_n=0$ which has the normalization
\beq
\label{fzero}
|f_{0R}(1)|^2=\frac{1-2\tilde{\nu}}{1-\epsilon^{1-2\tilde{\nu}}}.
\eeq
For $\tilde {\nu} \geq \frac{1}{2}$ this suppresses the wavefunction at the visible brane by the factor $\epsilon
^{\tilde{\nu}-\frac{1}{2}}$. On the other hand the higher KK bulk neutrino modes are not subjected to 
such a suppression. 
Similar to the bulk Higgs the KK bulk neutrinos have masses given by the equation
\beq
\label{kknm}
J_{\tilde{\nu}-\frac{1}{2}}(\tilde{x}_n)=0
\eeq
which can be seen [Eq.(\ref{newvari})] to be of order weak scale. Thus we expect no more than 
one or two such neutrino modes
are kinematically accessible to light Higgs boson decays. This is strikingly different from bulk neutrinos in
the factorizable geometry scenario \cite {fgnu} where a large number of neutrinos are available 
if the radius of compactification is sufficiently large. The phenomenology of the simplest model of this
type is discussed in \cite{nufg} and \cite {fgnu}.

        We now proceed to discuss the interaction between the brane leptons, the bulk Higgs bosons 
and the bulk neutrinos. The action is:
\beq
\label{bhf}
S_{b\nu}= - \sqrt{2} \, \hat{Y}_5\int d^4x \sqrt{g_{vis}}\, \bar{L}_0H_0(x,\pi)\Psi_R(x,\pi)+h.c.
\eeq
where $\hat{Y}_5$ is yet another Yukawa coupling.
After substituting in the KK decompositions of Eqs.(\ref{kkh}), (\ref{kkpsi}),
and Eqs. (\ref{newfv}) plus fields rescaling and  spontaneous
symmetry breaking we arrive at the couplings between $\nu_L$ and the KK bulk neutrino states.
These will provide the off diagonal terms for  neutrino mass matrix. Explicitly, they
are given by
\beq
\label{nmass}
\sum_{n}\hat{Y}_5 \int d^4x \, \epsilon v_0r_c\sqrt{v_0k}\, f_{nR}(\pi)\,\bar{\nu}_L\psi_{nR} + h.c.
\eeq 
For the bulk zero mode we obtain a mass term for $\nu_L$ with the value given by
\beq
\label{litenu}  
m_{\nu}=\hat{Y}_5v\sqrt{kr_c}\, \epsilon^{\tilde{\nu}-\frac{1}{2}}
\eeq
where we have used Eqs.(\ref{fzero}) and (\ref{fermi}).We shall demand that
$\mathrm {10^{-4}}$ eV $\leq m_{\nu} \leq $ 2 eV as indicated by current solar and atmospheric
neutrino data \cite{superk} as well as direct measurement of neutrino mass in tritium $\beta$ decays 
\cite {trit}. 
Assuming that $\hat{Y}_5$ is of order $1$ and using the values we have obtained previously on the parameters 
in Eq.(\ref{litenu})  
we find  that $\tilde{\nu}$ lies in the range 1.1 to 1.5 corresponding to 2 eV and $\mathrm {10^{-4}}$ eV light
neutrino respectively. As can be seen this conclusion is 
not sensitive to $\hat{Y}_5$ unless it is extremely large or small.
For the lower value of $\tilde {\nu}\sim $ 1.1 the roots of the half integer Bessel function 
is close to $j\pi$ where $j$ is
an integer. Thus, we find that diagonal terms of neutrino mass matrix of the KK neutrinos is well 
 approximated by
\beq
\label{kkmass}
\mu_n\simeq n\pi k \epsilon
\eeq
Similarly for $\tilde{\nu}\sim$ 1.5 the solutions of Bessel functions of order 1 apply. Both are independent
of $\hat{Y}_5$.  
It is interesting to note that the small value of $m_{\nu}$ is a result of the function $f_{1R}$ being
almost vanishing at $\phi=\pi$ \cite{GN} which in turn is due to the warp metric and the assumed separation
between the hidden and the visible branes.
 For details of the neutrino mass matrix see \cite {GN}.

        By an analogous calculation we derive  the effective coupling between $\nu_L$, the $m^{\mathrm th}$
bulk neutrino mode, and the $n^{\mathrm th}$ KK Higgs boson which we denote by $\tilde{y}_{nm}^{eff}$. Thus,
\beqa
\label{ynm}
\tilde{y}_{nm}^{eff} &=& \hat{Y}_5 y_n(\pi)f_{mR}(\pi)\sqrt{kr_c}e^{-kr_c\pi} \nonumber  \\ 
                     &=& \hat{Y}_5 kr_c\sqrt{\frac{2}{1-(\frac{\omega}{x_{n\nu}})^2}}
\eeqa
where we used the normalized values of $y_n$ and $f_{nR}$. We note that this effective coupling is not
small and universal for each of  the KK Higgs excitations; however, it does vary from one Higgs excitation
to another even though numerically this variation is not large.  To get
an idea of the numerics we take the lightest Higgs boson, $h_1$, and $\omega$=1. This results in
 $\tilde{y}^{eff}_1
=\mathrm {17.5}\hat{Y}_5$ and $\tilde{y}^{eff}_2=\mathrm {17.1}\hat{Y}_5$. The enhancement over the naive
Yukawa coupling is due to the fact that 
the wavefunctions of neither the bulk Higgs nor the neutrino states are  small on
the visible brane.  
Without resorting to 
fine tuning of the Yukawa coupling this leads to a large invisible width of the $h_1$
 if the channel is kinematically open.  For example, the width of $h_1$ into $\bar{\nu}_L \psi_{1R}$
is given by
\beq
\label{hnuw}
\Gamma_{\nu\psi_1}=\Gamma(h_1\rightarrow \bar{\nu}_L \psi_{1R} + \bar{\psi}_{1R}\nu_L) = \frac{m_1}{8\pi} 
\frac{(\hat{Y}_5 kr_c)^2}{1-\frac{\omega^2}{x_{1\nu}^2}} \left( 1-\frac{\mu_1^2}{m_1^2} \right)^2
\tan^2 \theta_{\nu} ,
\eeq
where $\tan \theta_{\nu}$ is the mixing between the lightest neutrino and its KK excitation. In principle
this can be constrained by the invisible width of the Z boson if the the KK neutrino is lighter then 90 GeV.
For heavier neutrinos we have to rely on details of the model, instead we shall take this to be a free
parameter.
 For $\hat{Y}_5=\frac{1}{3}$, $\omega=$ 1 
and $\tan \theta_{\nu} \sim $ 0.1, a 
150 GeV Higgs boson will have an 
invisible width, $\Gamma_{inv}=$170 MeV. This is to be compared with the b$\bar{\rm b}$ width, $\Gamma_b$ 
which we calculated
to be 32.8 MeV using Eq.(\ref{hee}) and the SM width, $\Gamma ^{SM}_b=$ 2.6 MeV. We have also
used the running $b$-quark mass which is the dominant QCD correction. 
 We also note the following interesting branching ratio:
\beq
\label{binv}
\frac{\Gamma_{\nu\psi_1}}{\Gamma_b}=\frac{1}{3\sqrt{2}} \frac{kr_c}{G_F}\left(\frac{\hat{Y}_5}{m_b} \right)^2
 \left(1-\frac{\mu_1^2}{m_1^2} \right)^2\left( 1-\frac{4m_b^2}{m_1^2}\right) ^{-\frac{3}{2}}\tan^2\theta_{\nu}
\eeq
        In Fig. 1 we display the branching ratio into $\mathrm b \bar{\mathrm b}$ of $h_1$ for the case  
with bulk neutrinos as a function of the mass $m_1$. The parameters used are given previously.  
The corresponding SM Higgs branching ratio is
also given as a comparison. One striking feature is the invisible width which is large for the values of
$\hat {Y_5}$ and $\tan {\theta_{\nu}}$ we used and this suppresses the branching ratio. In Fig.2 the
total decay width of the $h_1$ is plotted for different values of $m_1$. The enhancement factors are
evident.

        In conclusion we have constructed a model of bulk Higgs boson in the Randall-Sundrum scenario 
where the SM chiral fermions and gauge bosons are confined on the visible brane. By identifying the first
KK excitation of the 5D scalar as the lowest mass Higgs we find that the partial widths and hence 
the total decay width of such a boson are
enhanced compared to the SM. This enhancement is directly proportional to the quantity $kr_c$. This model
also predicts that the effective Higgs to $t\bar{t}$ is large which does not necessarily mean that 
perturbation is not applicable since the Yukawa coupling itself can still be of order unity. A detail
investigation of this issue is certainly worthwhile but is beyond the scope of the present paper.
 
        Furthermore, the phenomenology of the bulk Higgs changes drastically if we add bulk neutrinos    
into the picture. This has the added motivation of providing a mechanism for generating 
a small mass for the active $\nu_L$. In this case, the
invisible width of the Higgs boson is not negligible even though only one or two KK bulk neutrino decays are
open for Higgs masses between 125 to 250 GeV. Obviously, this has important consequences for Higgs boson
searches in high energy colliders. Other characteristics of the model include the mass spectra of the Higgs
and bulk neutrino which are given by roots of Bessel functions of various orders. Surprisingly, for the
bulk neutrinos for the phenomenologically interesting cases the order of the Bessel function varies between 
 1/2 to 1. This
makes the model much more predictive then naively expected. We eagerly await a rich phenomenology 
of the model discernibly different
from the SM to be discovered.
\newpage  
        {\bf Acknowledgement.} This work is partially supported by the Natural Science and
Engineering Council of Canada and the National Science 
Council of Taiwan, R.O.C (Grant No. 89-2112-M-003-003).  
C.H.C. would like to thank Darwin Chang, Chopin Soo, Miao Li, Y. C. Kao and P.M. Ho for useful discussions.
We thank
the National Center for Theoretical Sciences
of Taiwan, where most of the work was done, for hospitality and support. 
 
\newpage
\begin{table}
\label{tab1}
\caption{ The mass  of the first KK excitation,$m_2$ and values of $k\epsilon$ in parenthesis 
for different values of $\omega$ (first column) and $m_1$ in GeV(first row)}.
\begin{center}
\begin{tabular}{|c|c|c|c|c|}
\hline
\hspace*{-6pt}%
\raisebox{3.5mm}[0pt][0pt]{\begin{turn}{-29}\rule{20mm}{0.4pt}\end{turn}}%
\hspace*{-6pt}
& & & & \\[-2.8mm]

$\omega$ \hspace*{2mm} \raisebox{4mm}{$m_1$} &
125       &150        &200           &250  \\
\hline
0.1   &228 (32.6) &275 (39.1) &366 (52.1)   &458 (65.2)  \\
\hline
1.0   &224 (30.6) &266 (36.7) &358 (48.9)   &448 (61.1)   \\
\hline
2.0   &213 (26.4) &255 (31.7) &340 (42.3)   &425 (52.8)  \\
\hline
10.0  &169 (10.0) &203 (12.0) &270 (16.0)   &338 (20.0) \\
\hline
\end{tabular}
\end{center}
\end{table}

\clearpage
\newpage
\begin{figure}
  \centerline{\epsffile{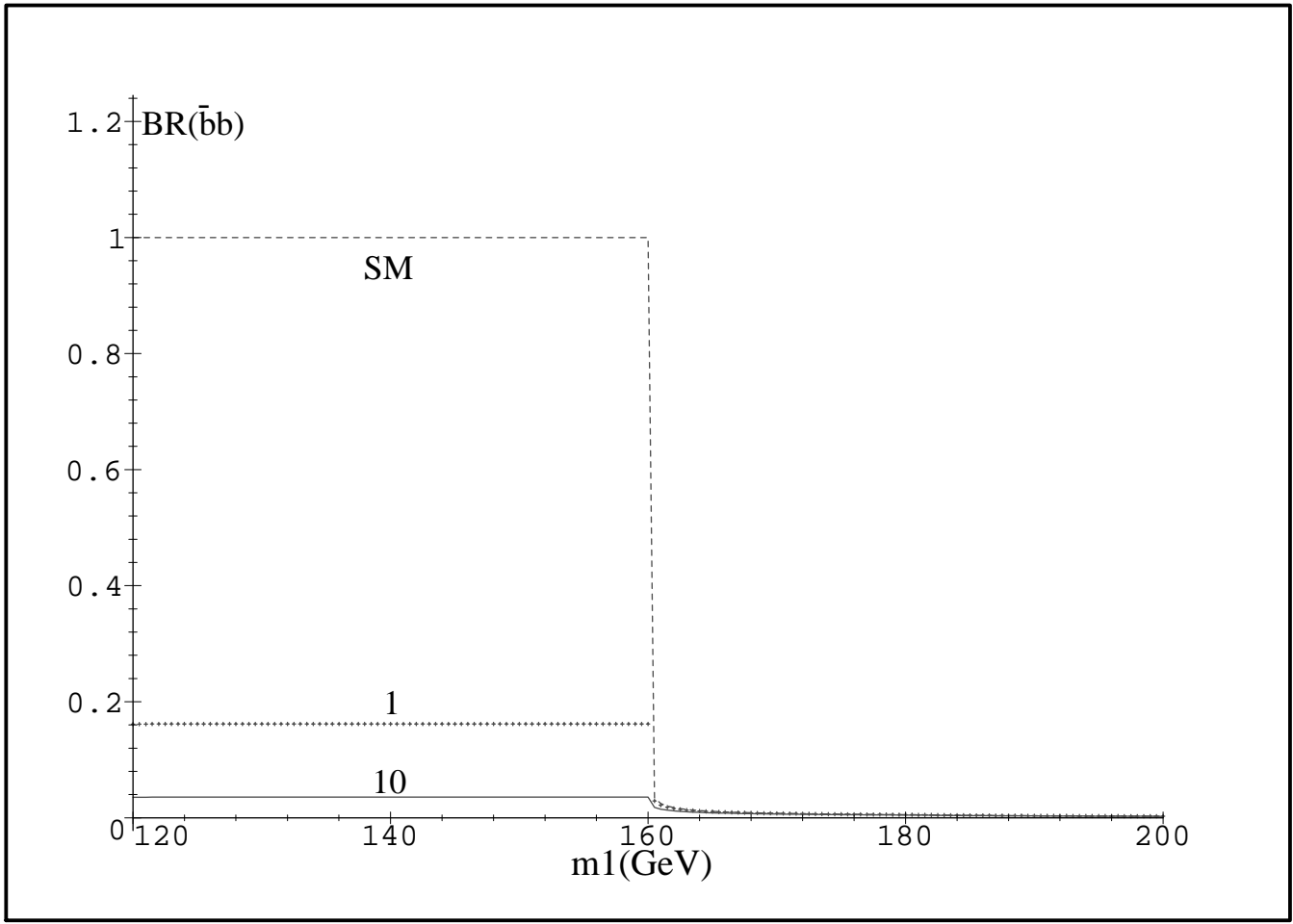}}
  \caption{\label{Fig1}
 The branching of the lowest mass Higgs into b$\bar{\mathrm b}$ as a function of 
the mass $m_1$. The values for $\omega=$1 and 10 are shown. The SM Higgs result is given by the dashed curve. }
\end{figure}

\newpage
\begin{figure}
\centerline{\epsffile{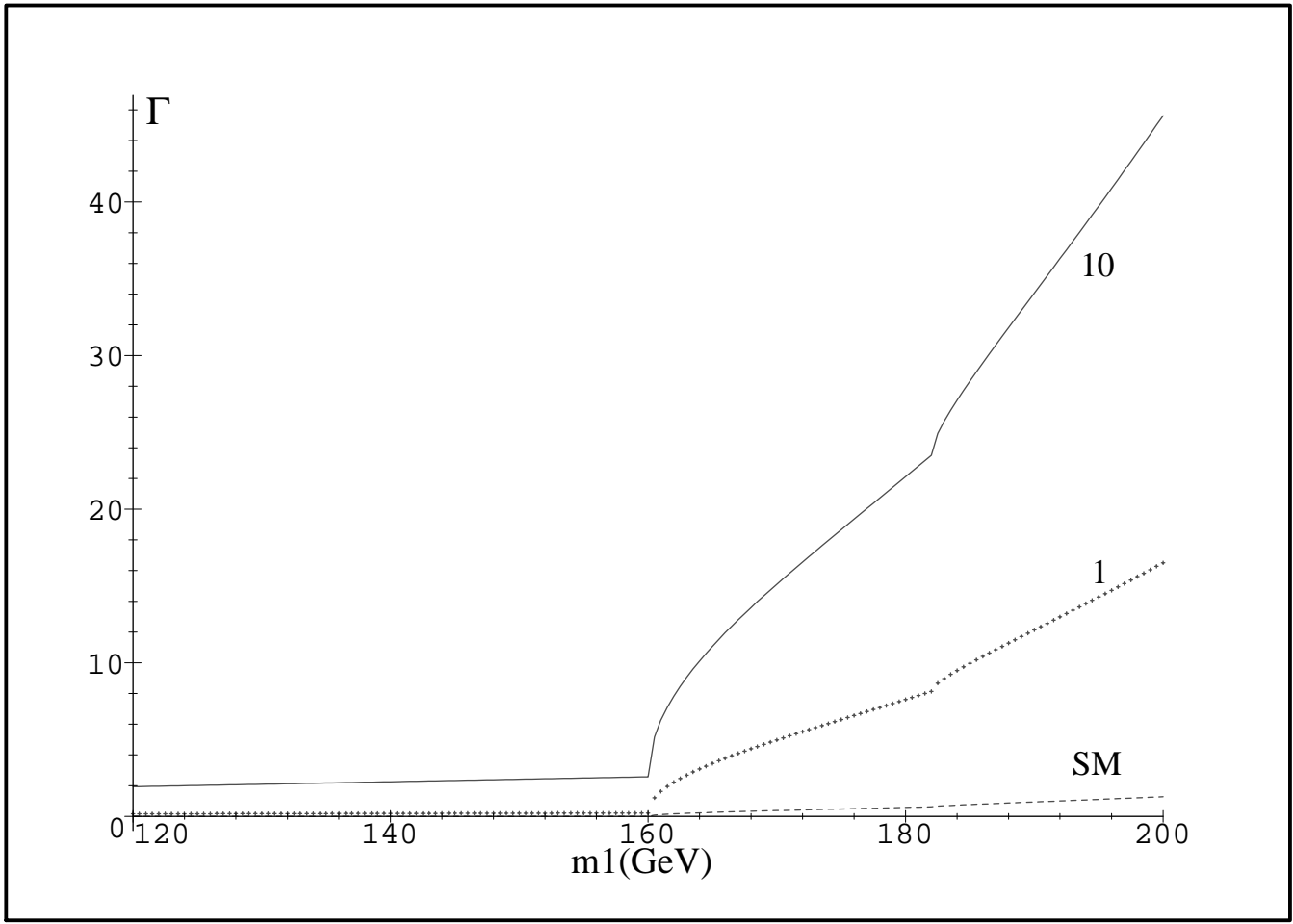}}
\caption{\label{Fig2}
 The total decay width of $h_1$ as a function of $m_1$. The width for the SM Higgs boson is given by the dashed
line}
\end{figure}
\newpage
%\setcounter{figure}{0}
%\begin{figure}
%\epsfsize=14cm
%\centerline{\epsffile{bhfig1.ps}}
%\end{figure}

%\begin{figure}
%\epsfsize=14cm
%\centerline{\epsffile{bhfig2.ps}}
%\end{figure}

\end{document}